\documentstyle[12pt,epsfig]{article}
\def\R{{\rm I\!R}}

\def\N{{\rm I\!N}}

\def\Q{{\mathchoice
 {\setbox0=\hbox{$\displaystyle\rm Q$}\hbox{\raise 0.15\ht0\hbox to0pt
 {\kern0.4\wd0\vrule height0.8\ht0\hss}\box0}}
 {\setbox0=\hbox{$\textstyle\rm Q$}\hbox{\raise 0.15\ht0\hbox to0pt
 {\kern0.4\wd0\vrule height0.8\ht0\hss}\box0}}
 {\setbox0=\hbox{$\scriptstyle\rm Q$}\hbox{\raise 0.15\ht0\hbox to0pt
 {\kern0.4\wd0\vrule height0.7\ht0\hss}\box0}}
 {\setbox0=\hbox{$\scriptscriptstyle\rm Q$}
 \hbox{\raise 0.15\ht0\hbox to0pt
 {\kern0.4\wd0\vrule height0.7\ht0\hss}\box0}}}}
\def\C{{\mathchoice
 {\setbox0=\hbox{$\displaystyle\rm C$}\hbox{\hbox to0pt
 {\kern0.4\wd0\vrule height0.9\ht0\hss}\box0}}
 {\setbox0=\hbox{$\textstyle\rm C$}\hbox{\hbox to0pt
 {\kern0.4\wd0\vrule height0.9\ht0\hss}\box0}}
 {\setbox0=\hbox{$\scriptstyle\rm C$}\hbox{\hbox to0pt
 {\kern0.4\wd0\vrule height0.9\ht0\hss}\box0}}
 {\setbox0=\hbox{$\scriptscriptstyle\rm C$}\hbox{\hbox to0pt
 {\kern0.4\wd0\vrule height0.9\ht0\hss}\box0}}}}
\font\fivesans=cmr5
\font\sevensans=cmr7
\font\tensans=cmr10
\newfam\sansfam
\textfont\sansfam=\tensans\scriptfont\sansfam=
 \sevensans\scriptscriptfont
\sansfam=\fivesans
\def\sans{\fam\sansfam\tensans}
\def\Z{{\mathchoice
 {\hbox{$\sans\textstyle Z\kern-0.4em Z$}}
 {\hbox{$\sans\textstyle Z\kern-0.4em Z$}}
 {\hbox{$\sans\scriptstyle Z\kern-0.3em Z$}}
 {\hbox{$\sans\scriptscriptstyle Z\kern-0.2em Z$}}}}
\newcommand{\beq}{\begin{equation}}
\newcommand{\eeq}{\end{equation}}
\newcommand{\bea}{\begin{eqnarray}}
\newcommand{\eea}{\end{eqnarray}}

\begin{document}
\begin{titlepage}
\begin{center}
{\Large
Perfect observables for the hierarchical \\[2mm]
non-linear $O(N)$-invariant $\sigma$-model
} \\[10mm]
\end{center}
\begin{center}
{\Large C. Wieczerkowski $^1$ and Y. Xylander $^2$} \\[10mm]
\end{center}
\begin{center}
$^1$ Institut f\"ur Theoretische Physik I,
Universit\"at M\"unster, \\
Wilhelm-Klemm-Stra\ss e 9, D-48149 M\"unster, \\
wieczer@yukawa.uni-muenster.de \\[2mm]
$^2$ Institut f\"ur Theoretische Physik II,
Universit\"at Hamburg, \\
Notkestra\ss e 85, D-22607 Hamburg \\
xylander@x4u2.desy.de
\end{center}
\vspace{-10cm}
\hfill
\begin{tabular}{r}
hep-lat/9505012\\
DESY-95-094 \\
MS-TP1-95-1
\end{tabular}
\vspace{11cm}
\begin{abstract}
We compute moving eigenvalues and the eigenvectors of
the linear renormalization group transformation for
observables along the renormalized trajectory of the
hierarchical non-linear $O(N)$-invariant $\sigma$-model
by means of perturbation theory in the running coupling
constant. Moving eigenvectors are defined as solutions
to a Callan-Symanzik type equation.
\end{abstract}
\end{titlepage}
\section{Introduction}
Hasenfratz and Niedermayer calculated in \cite{HN} the
asymptotic form of the renormalized trajectory called
perfect action of the two dimensional $O(N)$-model
with non-linear block spin. They observed that the
block spin transformation could be treated classically
in the asymptotically free domain to an excellent
approximation. In \cite{WX} we introduced an improvement
scheme in which the zeroth order approximation is the
perfect action. This scheme is a combination of
perturbation theory in the running coupling with the
idea of scaling. We worked it out for the $O(N)$-model
with linear block spin in the hierarchical approximation.
The improved action is not yet the complete improved
theory. The complete improved theory should also
include improved observables. In this paper we
show how the improvement program in \cite{WX} can be
extended to observables for the hierarchical $O(N)$-model.

In Wilson's renormalization group \cite{WK} a Euclidean
quantum field theory is thought of in terms of the flow
of effective interactions $V(\phi)$ generated by a block
spin transformation ${\cal R}$. In the following $\phi$
will denote an $N$-component real valued field on a two
dimensional unit lattice. The interactions will be
invariant under global $O(N)$ transformations. ${\cal R}$
will respect global $O(N)$ invariance.
The fixed points ${\cal R}V_*=V_*$ of ${\cal R}$ describe
scale invariant (continuum) field theories. The
eigenvalues of the linearized block spin transformation
at a fixed point determine the associated critical
exponents. The two dimensional $O(N)$-invariant non-linear
$\sigma$-model is expected to possess only two fixed points
for $N>2$ (and $N>1$ in the hierarchical approximation),
an asymptotic, unstable ultraviolet fixed point
$V_{UV}$ and a stable infrared fixed point $V_{IR}$.
The model is asymptotically free in the ultraviolet:
$V_{UV}$ is a theory of $N-1$ massless free fields.
In the hierarchical case $V_{UV}$ is better thought of from
the orthogonal point of view as a high temperature fixed
point of a single component radial field.
The fixed points are expected to be connected by a one
dimensional renormalized trajectory which is stable under
block spin transformations. The points on the
renormalized trajectory describe (continuum) theories
with broken scale invariance. Interactions on the
renormalized trajectory are said to scale.
The renormalized trajectory is both of principal and
of practical interest since it admits the description of
a true continuum theory (not a lattice approximation)
in terms of a lattice theory. It is the ultimate
improved action in the sense of Symanzik \cite{S}.

Let the renormalized trajectory be parametrized in terms
of a local coordinate $f$. Recall that it is a curve in
the space of $O(N)$ invariant interactions. The block spin
transformation then becomes
\beq
{\cal R}V_{RT}(\phi|f)=V_{RT}(\phi|\beta(f)).
\label{CS}
\eeq
A natural local coordinate in the vicinity of the
asymptotically free fixed point is the running coupling.
The running coupling will be defined as the leading
trilinear $\sigma$-$\pi$-$\pi$-vertex. In the hierarchical
approximation it is identical with the inverse radius.
The complete dynamics of the renormalization group
on the renormalized trajectory is encoded in the
flow of the running coupling. We call $\beta(f)$
Callan-Symanzik function since it defines this
discrete flow.

A block spin transformation comes together with
a linear mapping of observables which preserves their
expectation values. This mapping is the linearized
renormalization group transformation. We shall consider
this linear mapping ${\cal LR}$ to the block spin
transformation over the renormalized trajectory.
An observable will be called an eigenvector with
eigenvalue $\epsilon(f)$ if it satisfies the equation
\beq
{\cal L}_{RT}{\cal R O}(\phi|f)=
\epsilon(\beta(f)){\cal O}(\phi|\beta(f)).
\label{CSO}
\eeq
In accordance with the terminology for the interaction
we will speak of this property as scaling.
The space of eigenvectors defines a moving frame in
the tangent space over the renormalized trajectory. The
set of eigenvalues will be called the spectrum of the theory.
It is exactly calculable at the fixed point. Moving away from
the fixed point on the renormalized trajectory it becomes
perturbed. In the vicinity of the ultraviolet fixed point it is
therefore natural to perform a perturbation expansion in the
running coupling to determine the spectrum from (\ref{CSO}).
In this paper we address the question of computing
these perturbative corrections. At the fixed point (\ref{CSO})
becomes the eigenvalue equation for the critical exponents.
Away from the fixed point the spectrum governs the
rate of attraction towards the renormalized trajectory.
The set of eigenvectors forms a basis of the linear space
of observables. Associated with this basis is a set of
fusion rules. We will give a general expression for
correlation functions of eigenvectors on the renormalized
trajectory in terms of the spectrum and the fusion rules.
The investigation of the spectrum and the moving frame
complete the analysis begun in \cite{WX}.

We should mention that the two dimensional hierarchical
$O(N)$-model has been rigorously studied by Gawedzki and
Kupiainen in \cite{GK}, where both the existence of an
ultraviolet limit and the existence of a renormalized
trajectory has been established beyond perturbation theory.
This result has also been obtained by Pordt and Reisz in
\cite{PR} using a slightly different technology. Both
groups however do not address the question of the spectrum.
A rigorous study of the full $O(N)$-model both in the
ultraviolet and more importantly in the infrared domain
is still an outstanding problem.

\section{Hierarchical $O(N)$ model}
Let us consider the renormalization group flow generated
by the hierarchical block spin transformation\footnote{
Note that (\ref{hrgt}) differs from the transformation
used in \cite{WX}. We have exchanged the integration with
the rescaling step.}
\beq
e^{-{\cal R}V(\psi)}=
{\cal N}\left(
\int{\rm d}\mu_\gamma(\zeta)
e^{-V(\psi+\zeta)}\right)^2.
\label{hrgt}
\eeq
in $D=2$ dimensions with scale parameter $L=\sqrt{2}$.
The interaction $V(\phi)$ is taken to be a function
of a real $N$-component variable $\phi$.
\beq
{\rm d}\mu_\gamma(\zeta)=
(2\pi\gamma)^{\frac{-N}2}
e^{\frac{-\zeta^2}{2\gamma}}{\rm d}^N\zeta
\eeq
is the Gaussian measure on $\R^N$ with mean zero and
covariance ${\rm diag}(\gamma,\dots,\gamma)$. The subspace
of $O(N)$ invariant interactions is stable under (\ref{hrgt}).
We will restrict our attention to this subspace. $O(N)$
invariance requires $V(\phi)$ to be a function of the modulus
$|\phi|$. The normalization constant in (\ref{hrgt}) is
conveniently chosen such that the interaction is always zero
at its minimum. We will denote this minimum by $r$ and call it
the radius of the potential.

Let us define the renormalized trajectory as the curve
$V_{RT}(\phi|f)$ in the space of $O(N)$ invariant potentials,
parametrized by the inverse radius $f=\frac1r$, with the
following two properties:\\[2mm]
1) $V_{RT}(\phi|f)$ is stable under the block spin transformation
${\cal R}$. It follows that there exists a function $\beta(f)$
such that
\beq
{\cal R}V_{RT}(\phi|f)=V_{RT}(\phi|\beta(f)).
\label{con1}
\eeq
In other words a block spin transformation acts on the
renormalized trajectory by a transformation of the coordinate
given by a $\beta$-function.\\[2mm]
2) The asymptotic behavior of $V_{RT}(\phi|f)$ as the running
coupling $f$ goes to zero is given by
\bea
V_{RT}(\phi|f)&=&V^{(1)}_{RT}(\phi|f)+O(f^2), \nonumber\\
V^{(1)}_{RT}(\phi|f)&=&\frac1{2\gamma}
\left(|\phi|-\frac1f \right)^2.
\label{con2}
\eea
Up to corrections of second order in the running coupling
the renormalized trajectory coincides with the perfect action.
\\[2mm]
At least in terms of perturbation theory in the running coupling
(\ref{con1}) and (\ref{con2}) have a unique solution as has been
shown in \cite{WX}. Let us remark that (\ref{con2}) is compatible
with (\ref{con1}) since it reproduces itself under (\ref{hrgt})
according to
\bea
{\cal R}V^{(1)}_{RT}(\phi|f)&=&
V^{(1)}_{RT}(\phi|\beta(f))+O(f^2),\nonumber\\
\beta(f)&=&f+O(f^2).
\eea
up to second order corrections. The general form of the solution
to (\ref{con1}) and (\ref{con2}) in terms of perturbation theory
is
\bea
V_{RT}(\phi|f)&=&\sum_{n=2}^\infty
P_n(f)\left(|\phi|-\frac1f\right)^n,
\nonumber\\
P_2(f)&=&\frac1{2\gamma}+\sum_{m=0}^\infty
c_{2,2m+2}f^{2m+2},
\nonumber\\
P_n(f)&=&\sum_{m=0}^\infty c_{n,2m+n} f^{2n+m},
\nonumber\\
\beta(f)&=&f+\sum_{m=0}^\infty b_{2m+3} f^{2m+3}.
\label{rtra}
\eea
The coefficients up to fifth order in the running coupling
turn out to be given by
\bea
c_{2,2}&=&-\frac{3(N-1)}4,\nonumber\\
c_{2,4}&=&\left(\frac{9(N-1)^2}8-\frac{61(N-1)}{28}\right)
\gamma,\nonumber\\
c_{3,3}&=&\frac{7(N-1)}{18},\nonumber\\
c_{3,5}&=&\left(-\frac{5(N-1)^2}8+\frac{257(N-1)}{180}\right)
\gamma,\nonumber\\
c_{4,4}&=&-\frac{15(N-1)}{56},\nonumber\\
c_{5,5}&=&\frac{31(N-1)}{150},\nonumber\\
b_3&=&\frac{(N-1)\gamma}2,\nonumber\\
b_5&=&\left(\frac{3(N-1)^2}8+\frac{13(N-1)}{12}\right)
\gamma^2,\nonumber\\
\eea
It follows that the model is asymptotically free for $N>1$.
It is custom to use $g=f^2$ instead of $f$ as running coupling
in discussions on asymptotic freedom. From the $\beta$-function
to fifth order in $f$ it follows that the recursion relation
to third order in $g$ is given by
\bea
\overline{\beta}(g)&=&g+\sum_{m=2}^\infty
\overline{b}_m g^m,\nonumber\\
\overline{b}_2&=&(N-1)\gamma,\nonumber\\
\overline{b}_3&=&\left((N-1)^2+\frac{13(N-1)}6\right)
\gamma^2.
\label{free}
\eea
Thus $g$ is marginally relevant for $N>1$.

This definition of a renormalized trajectory does not refer to
a continuum limit procedure. It is nevertheless identical with
the continuum limit effective potential of models in the
$O(N)$ universality class in the hierarchical renormalization
scheme. To prove this we can perform the continuum limit using
(\ref{rtra}) as bare interaction. Define the bare coupling to
be the $n$-fold preimage $\beta^{-n}(f)$ of a renormalized
value $f$. By construction of the renormalized trajectory it
follows that
\beq
{\cal R}^n V_{RT}(\phi|\beta^{-n}(f))=
V_{RT}(\phi|f)
\label{bare}
\eeq
for all numbers $n$ of renormalization group steps. The
continuum limit $n\rightarrow\infty$ is immediately performed
since the right hand side of (\ref{bare}) is independent of
$n$. All that is needed is an analysis of the recursion
relation defined by the $\beta$-function or rather its inverse.
This is a comparatively easy task. See for instance \cite{GK}.
For $N>1$ it follows from (\ref{free}) that the bare coupling
tends to zero as $n$ goes to infinity. (One needs to take into
account logarithmic corrections piled up by the term of third
order in $g$, fifth order in $f$.) It follows that the
perturbation expansion in the running coupling is valid.
To make contact with the hierarchical real world one should
also assign a scale, for instance in form of a lattice
spacing $a$, to the point on the renormalized trajectory
where the running coupling is given by the renormalized value
$f$. The bare cutoff in (\ref{bare}) is then $L^{-n}a$ with
$L$ the block scale.

The sceptical reader may worry to what extent this construction
is connected with the continuum limit of a bare theory defined
by his favorite $O(N)$ invariant interaction. Consider for
instance the standard interaction of the linear $O(N)$-model
defined by
\beq
V(\phi|f) = \lambda f^2
\left(\phi^2-\frac1{f^2}\right)^2.
\label{stan0}
\eeq
Its continuum limit is constructed as the result of the infinite
iteration
\beq
\lim_{n\rightarrow\infty}{\cal R}^n
V(\phi|f_{-n}(f))=V_{cont}(\phi|f).
\label{stan1}
\eeq
Here the bare coupling $f_{-n}(f)$ is tuned such that the minimum
of ${\cal R}^n V(\phi|f_{-n}(f))$ is located at the renomalized
radius $|\phi|=\frac1f$. The continuum limit is universal as we
have learnt from the work of Wilson \cite{WK}. Therefore the
connection to the above definition is simply
\beq
V_{cont}(\phi|f)=V_{RT}(\phi|f).
\label{stan2}
\eeq
A rigorous proof of the existence of (\ref{stan1}) and, inbetween
the lines, also of (\ref{stan2}) has been given by Gawedzki and
Kupiainen \cite{GK}. More generally, the set of all bare interactions
sharing (\ref{rtra}) as their common continuum limit defines the
universality class of the hierarchical $O(N)$-model.
(\ref{stan0}) is known to belong to this class and so is the original
model with sharp constraint. So what we do, when we define the
renormalized trajectory by the above two conditions, is to completely
disentangle the admittingly also interesting question if, how, and
at what pace the continuum limit is reached by some particular bare
model.

In practice we may not be able to compute (\ref{rtra}) to all
orders of perturbation theory in the running coupling for the
models we are really interested in, for example the full
nonlinear $O(N)$-model in terms of block spin transformation on
a unit lattice. In this situation the best one can do is to take
the highest order approximation to (\ref{rtra}) accessible as
bare interaction. In the hierarchical model for instance already
the second approximant
\beq
V^{(2)}_{RT}(\phi|f)=
\left(\frac1{2\gamma}-\frac{3(N-1)}4f^2\right)
\left(|\phi|-\frac1f\right)^2
\eeq
turns out to be an excellent starting point for a numerical
study of the renormalized trajectory. Its main property is
of course
\beq
{\cal R}V^{(2)}_{RT}(\phi|f)=
V^{(2)}_{RT}(\phi|\beta(f))+O(\beta(f)^3).
\eeq
It therefore coincides with the renormalized trajectory up to
corrections of third order in $f$. Let us finally mention that
the third order approximant would be ideally suited for a
rigorous construction along the lines of Gawedzki and Kupiainen
\cite{GK} and of Pordt and Reisz \cite{PR}.
The perturbative part is trivial since the
action reproduces itself to third order. The construction
therefore reduces to the proof of a stability bound which
controls non perturbative corrections. This bound is
already implicitly contained in the rigorous work on the
$O(N)$-model with standard bare interaction (\ref{stan0}).

\section{Perfect Observables}
Let us consider the hierarchical renormalization group
transformation for local observables corresponding to
(\ref{hrgt}) in the case of $D=2$ dimensions with scale
parameter $L=\sqrt{2}$. It is given by the linear
transformation
\beq
{\cal L}_V{\cal RO}(\psi)=
\frac{\int{\rm d}\mu_\gamma(\zeta)
e^{-V(\psi+\zeta)}{\cal O}(\psi+\zeta)}
{\int{\rm d}\mu_\gamma(\zeta)
e^{-V(\psi+\zeta)}}.
\label{lobx}
\eeq
The block volume is $L^D=2$. (\ref{lobx}) is the linearization
of (\ref{hrgt}) divided by the block volume
\beq
{\cal L}_V{\cal RO}(\psi)=
\frac12\frac{\partial}{\partial z}
{\cal R}(V+z{\cal O})\vert_{z=0}.
\eeq
In other words the transformation of observables is the tangent
map of the transformation of the potential.

The setup of perturbation theory for the linear transformation
(\ref{lobx}) is as follows. Since the model is $O(N)$-invariant
we can choose the block spin to be given by $\psi=
(r+\Psi)\hat{e}$ with $\hat{e}$ any $N$ component unit vector.
That is, we can trade $\psi$ for $\Psi=|\psi|-r$.
Observables will also be assumed $O(N)$-invariant. The
shift by $r$ serves to place us into the minimum of the potential.
We then decompose orthogonally the fluctuation field into
a radial and a tangential part with respect to the direction of
the block spin $\psi$. The decomposition is $\zeta=
\sigma\hat{e}+\pi$. The one component variable $\sigma$ is the
radial fluctuation field. The $(N-1)$-component variable
$\pi$ is the tangential fluctuation field. The Gaussian measure
factorizes into
\beq
{\rm d}\mu_\gamma(\zeta)=
{\rm d}\mu_\gamma(\sigma)\,
{\rm d}\mu_\gamma(\pi).
\eeq
We then insert this decomposition of $\zeta$ in the formula
for the potential on the renormalized trajectory and extract
the terms of order zero in the running coupling. The result
is
\beq
V_{RT}((r+\Psi+\sigma)\hat{e}+\pi|f)=
\frac{(\Psi+\sigma)^2}{2\gamma}+
V_{RT}^{(1)}(\Psi+\sigma,\pi|f).
\eeq
The zeroth order terms cannot be treated as perturbations. They
are fortunately only linear and quadratic in the radial
fluctuation field. They and can therefore be accomodated for
using the identity
\beq
{\rm d}\mu_\gamma(\sigma)
e^{\frac{-(\Psi+\sigma)^2}{2\gamma}}=
{\rm d}\mu_{\frac\gamma 2}(\sigma+\frac\Psi 2)
\,\frac1{\sqrt{2}}\, e^{\frac{-\Psi^2}{4\gamma}}.
\eeq
We then shift the radial fluctuation field to $\xi=
\sigma+\frac\Psi 2$. After these standard manipulations
the transformation (\ref{lobx}) becomes
\beq
{\cal L}_{RT}{RO}((r+\Psi)\hat{e})=
\frac{\int{\rm d}\mu_{\frac\gamma 2}(\xi)
\int{\rm d}\mu_\gamma(\pi)
e^{-V_{RT}^{(1)}(\frac\Psi 2+\xi,\pi|f)}
{\cal O}((r+\frac\Psi 2+\xi)\hat{e}+\pi)}
{\int{\rm d}\mu_{\frac\gamma 2}(\xi)
\int{\rm d}\mu_\gamma(\pi)
e^{-V_{RT}^{(1)}(\frac\Psi 2+\xi,\pi|f)}}.
\eeq
At this point perturbation theory is applicable. The expansion
parameter is the inverse radius $f=\frac1r$. For the observables
we have in mind we also have to perform an expansion in terms
of the running coupling.

An observable is called a moving eigenvector to the moving
eigenvalue $\epsilon(f)$ if it satisfies the renormalization
group equation
\beq
{\cal L}_{RT}{\cal RO}(\phi|f)=
\epsilon(\beta(f)) {\cal O}(\phi|\beta(f)).
\label{osca}
\eeq
Solutions to (\ref{osca}) are perfect observables. A
parametrization well suited for perturbation theory is
\beq
{\cal O}(\phi|f)=
\sum_{n=0}^\infty Q_n(f)
\left(|\phi|-\frac1f\right)^n.
\label{para}
\eeq
Here the coefficients $Q_n(f)$ are taken to be power
series in the running coupling $f$. (The coefficients are
not expected to exhibit any singularity at zero coupling.)
We will organize the solutions of (\ref{osca})
according to their zeroth order coefficients. The
zeroth order observables are simply normal ordered
monomials. Let us also perform the perturbation expansion
for the observables of the form (\ref{para}).
Define
\beq
W(\phi|f)=|\phi|-\frac1f.
\eeq
As in the potential we find a term of order zero
in the running coupling. Separating it off we obtain
\beq
W((\frac1f+\frac\Psi 2+\xi)\hat{e}+\pi|f)=
\frac\Psi 2 +\xi+
W^{(1)}(\frac\Psi 2+\xi,\pi|f).
\eeq
In the limit when the radius becomes infinite the transformation
(\ref{lobx}) reduces to the convolution with the radial
Gaussian measure. This transformation is identical with that
of a one component scalar field in two dimensions at the
high temperature fixed point. We are therefore immediately lead
to normal ordered monomials. Put $\gamma^\prime=\frac{2\gamma}3$.
Then it follows that
\beq
\int{\rm d}\mu_{\frac\gamma 2}(\xi)
:(\frac\Psi 2+\xi)^n:_{\gamma^\prime}=
\frac1{2^n} :\Psi^n:_{\gamma^\prime}.
\eeq
Therefore the eigenvectors are normal ordered monomials to
zeroth order as expected. The zeroth order spectrum is
$\epsilon_n=\frac1{2^n}$. Let us write the normal ordered
monomials in the form
\beq
:\Psi^n:_{\gamma^\prime}=
\sum_{m=0}^n h_{n,m} \Psi^m.
\eeq
They are of course rescaled Hermite polynomials in the
variable $\Psi$. We then write the associated sequence of
observables defined as solutions of the scaling equation
(\ref{osca}) in the form
\beq
{\cal O}_n(\phi|f)=
\sum_{m=0}^\infty Q_{n,m}(f)
\left(|\phi|-\frac1f\right)^m.
\eeq
The coefficients $Q_{n,m}(f)$ are given by the power
series expansions
\beq
Q_{n,m}(f)=\sum_{l=0}^\infty
d_{n,m}^{(l)} f^l
\label{powe}
\eeq
with zero order coefficients of the normal ordered form
\beq
Q_{n,m}^{(0)}=h_{n,m}.
\eeq
That is, the observables are perturbations of normal
ordered monomials. The perturbative form of the
moving spectrum is
\beq
\epsilon_n(f)=\sum_{m=0}^\infty \epsilon_n^{(m)}f^m
\label{sowe}
\eeq
with zero order coefficients
\beq
\epsilon_n^{(0)}=\frac1{2^n}.
\eeq
This completes the setup for the improvement program for
observables. Let us then turn to the question of how to
compute the higher coefficients in the expansions
(\ref{powe}) and (\ref{sowe}) for the moving eigenvectors
and observables. The strategy is an adaptation of the
improvement program for the potential. Let us choose the
first (nontrivial) observable ${\cal O}_1(\phi|f)$
as an example and perform a third order computation to
some detail. The zeroth order argument provides us with
the information that
\bea
{\cal O}_1(\phi|f)&=&W(\phi|f)+O(f), \\
\epsilon_1(f)&=&\frac12+O(f).
\eea
Let us denote the zeroth order approximation by
${\cal O}(\phi|f)=W(\phi|f)$. As we will see
this observable already scales to second order.
We immediately perform a perturbation expansion to
third order for the effective observable of the
zeroth order approximation. The result is
\bea
\left({\cal L}_{RT}{\cal R}\right){\cal O}(\phi|f)&=&
\epsilon(f^\prime)\left(Q_0(f^\prime)+W(\phi|f^\prime)
+Q_2(f^\prime)W(\phi|f^\prime)^2\right)+
O((f^\prime)^4),\nonumber\\
\epsilon(f^\prime)&=&\frac12+\frac{(N-1)\gamma}4
(f^\prime)^2,\nonumber\\
Q_0(f^\prime)&=&-\frac{5(N-1)\gamma^2}{3}
(f^\prime)^3,\nonumber\\
Q_2(f^\prime)&=&-\frac{(N-1)\gamma}6(f^\prime)^3,
\eea
where the old running coupling $f$ is expressed in terms
of the new running coupling $f^\prime=\beta(f)$.
(${\cal L}_{RT}{\cal R}$ is understood as an operator
which is applied to the function ${\cal O}(\phi|f)$.)
This change of coupling prepares in particular the ground
for further iterations. Let us emphasize that not only
the coefficients but also the coordinate functions depend
on the running coupling and have to be adjusted.
{}From this we conclude that the observable already scales
to second order. The eigenvalue is therefore already correct to
second order. To third order both a constant and a term
quadratic in $W(\phi|f^\prime)$ are generated. Therefore
the observable does not reproduce its dependence on
the field to third order. To find the observable which
scales to third order we make the ansatz
\beq
{\cal O}(\phi|f)=d_0f^3+
W(\phi|f)+d_2f^3W(\phi|f)^2.
\label{ansa}
\eeq
The ansatz involves two improvement parameters $d_0$ and
$d_2$. To determine their value one again computes the
effective observable starting from (\ref{ansa}). The
expansion gives
\bea
\left({\cal L}_{RT}{\cal R}\right){\cal O}(\phi|f)&=&
\epsilon(f^\prime)\left(Q_0(f^\prime)+W(\phi|f^\prime)
+Q_2(f^\prime)W(\phi|f^\prime)^2\right)+
O((f^\prime)^4),\nonumber\\
\epsilon(f^\prime)&=&\frac12+\frac{(N-1)\gamma}4
(f^\prime)^2,\nonumber\\
Q_0(f^\prime)&=&\left(-\frac{5(N-1)\gamma^2}{3}
+d_2\gamma+2d_0\right)
(f^\prime)^3,\nonumber\\
Q_2(f^\prime)&=&\left(-\frac{(N-1)\gamma}6
+\frac{d_2}2\right)(f^\prime)^3.
\eea
The improvement parameters are then determined by the
simple linear equations
\bea
d_2&=&\frac{-(N-1)\gamma}6+\frac{d_2}2,\nonumber\\
d_0&=&\frac{-5(N-1)\gamma^2}3+d_2\gamma+2d_0.
\eea
The solution is
\beq
d_2=\frac{-(N-1)\gamma}3,\quad
d_0=2(N-1)\gamma^2.
\eeq
To third order we therefore find
\beq
{\cal O}_1(\phi|f)=
2(N-1)\gamma^2f^3+W(\phi|f)-
\frac{(N-1)\gamma}3f^3W(\phi|f)^2+
O(f^4)
\eeq
together with the eigenvalue
\beq
\epsilon_1(f)=\frac12+\frac{(N-1)\gamma}4f^2+
O(f^4).
\eeq
Note that the eigenvalue does not have a term of
third order in the running coupling. This scheme is
iterated in the obvious manner. Suppose that we have
computed the observable to order $s$ in the running
coupling
\beq
{\cal O}_1(\phi|f)=
{\cal O}_1^{(s)}(\phi|f)+O(f^{s+1}).
\eeq
Suppose that the order $s$ improved observable is given
by the general form
\bea
{\cal O}_1^{(s)}(\phi|f)&=&
\sum_{n=0}^{s-1}Q_{1,n}^{(s)}(f) W(\phi|f)^n,
\nonumber\\
Q_{1,0}^{(s)}(f)&=&
\sum_{r=0}^s d_{1,0}^{(r)}f^r,
\nonumber\\
Q_{1,1}^{(s)}(f)&=&1,
\nonumber\\
Q_{1,n}^{(s)}(f)&=&
\sum_{r=n+1}^s d_{1,n}^{(r)}f^r.
\label{ords}
\eea
Then by induction it follows that the effective observable
to order $s+1$ is again of this general form. It follows
that to every order of perturbation theory only finitely
many powers of the coordinate functions $W(\phi|f)$
appear. The sums turn out to involve either even or odd
powers in the running coupling respectively. We assume
that ${\cal O}_1^{(s)}(\phi|f)$ scales to order $s$.
That is,
\beq
\left({\cal L}_{RT}{\cal R}\right){\cal O}_1^{(s)}(\phi|f)=
{\cal O}_1^{(s)}(\phi|f^\prime)+O((f^\prime)^{s+1}).
\eeq
Then we take an ansatz of the form (\ref{ords}) to order
$s+1$ treating the coefficients of order $s+1$ in the
running coupling as improvement parameters. We compute
the effective observable. It depends linearly on the
improvement parameters. To order $s+1$ it has no other
choice. Then we claim invariance to obtain a linear system
of equations for the improvement coefficients. This
system turns out always to have a unique solution: the
improved observable.

\section{Hierarchical fusion rules}
The outcome of our analysis is a system of observables
${\cal O}_n(\phi|f)$ on the renormalized trajectory
parametrized by $f$. Under a hierarchical renormalization
group transformation ${\cal O}_n(\phi|f)$ is multiplied
by the moving eigenvalue $\epsilon_n(\beta(f))$ and the
coordinate is changed to $\beta(f)$. This scheme is
obviously iteratable which is the reason why we introduced
it from the beginning. To compute general correlation
functions we need one more ingredience which is the notion
of hierarchical fusion rules. The general form of our
observables in terms of $\Psi=|\phi|-r$ is
\bea
{\cal O}_n(\phi|f)&=&
\sum_{r=0}^\infty {\cal O}_n^{(r)}(\Psi)\,f^r,
\nonumber \\
{\cal O}_n^{(0)}(\Psi)&=&\,:\Psi^n:,
\nonumber \\
{\cal O}_n^{(1)}(\Psi)&=&0,
\nonumber \\
{\cal O}_n^{(2)}(\Psi)&=&
\sum_{m=0}^{\left[\frac{n-2}2\right]}
{\cal O}_{n,n-2-2m}^{(2)}\,:\Psi^{n-2-2m}:,
\nonumber \\
{\cal O}_n^{(r)}(\Psi)&=&
\sum_{m=0}^{\left[\frac{n+r-2}2\right]}
{\cal O}_{n,n+r-2-2m}^{(r)}\,:\Psi^{n+r-2-2m}:.
\eea
The normal ordering covariance is $\gamma^\prime=\frac23\gamma$.
Let us put $\gamma=1$ to simplify the notation. To zeroth order
in $f$ we rediscover normal ordered monomials in $\Psi$.
Their perturbations along the renormalized trajectory prove
to have no first order terms in the running coupling $f$.
To every order of perturbation theory in $f$ we find only
finitely many normal ordered powers of $\Psi$. The highest
power is $n+r-2$ for $r>2$. Associated with this system of
observables is a system of fusion rules defined by
\bea
{\cal O}_n(\phi|f){\cal O}_m(\phi|f)
&=&\sum_{l=0}^\infty N_{n,m;l}(f){\cal O}_l(\phi|f),
\nonumber\\
N_{n,m;l}(f)&=&\sum_{r=0}^\infty N_{n,m,;l}^{(r)}\,f^r.
\eea
Furthermore from the fusion rules we obtain a symmetric
bilinear form on the linear space of observables. It is
defined by
\beq
\left({\cal O}_n(\phi|f),{\cal O}_m(\phi|f)\right)=
N_{n,m;0}(f).
\eeq
The physical significance of this bilinear form is that
in the thermodynamic limit only the overlap of an
observable with the constant term is expected to survive.
Since correlation functions are expected to decrease with
distance the spectrum consists of eigenvalues strictly
smaller than one on the renormalized trajectory. To zeroth
order of perturbation theory we recapture the fusion rules
of normal ordered products
\bea
N_{n,m;l}^{(0)}&=&
\frac{n! m!}{n^\prime! m^\prime! l^\prime!}
(\gamma^\prime)^{l^\prime},\nonumber\\
n^\prime&=&\frac12 (l+m-n),\nonumber\\
m^\prime&=&\frac12 (n+l-m),\nonumber\\
l^\prime&=&\frac12 (n+m-l),
\label{fus0}
\eea
for $|n-m|\leq l\leq n+m$ and $n+m-l\in 2\Z$, zero else.
Furthermore to zeroth order the observables are orthogonal
with respect to the bilinear form
\beq
N_{n,m;l}^{(0)}=\delta_{n,m}\, m!\, (\gamma^\prime)^m.
\label{bil0}
\eeq
The simple pattern (\ref{fus0}), (\ref{bil0}) becomes perturbed
as one moves away from the ultraviolet fixed point on the
renormalized trajectory. The perturbation expansion for
$N_{1,1;l}(f)$ to fifth order in $f$ is for instance
\bea
N_{1,1;0}(f)&=&
\frac23+\frac{5(N-1)}9\, f^2-
\left(\frac{20(N-1)^2}{27}-\frac{233(N-1)}{63}\right)\,
f^4+O(f^6),\nonumber\\
N_{1,1;1}(f)&=&-2(N-1)\,f^3+
\left(\frac{17(N-1)^2}3-26(N-1)\right)\,f^5+O(f^7),\nonumber\\
N_{1,1;2}(f)&=&1+O(f^6),\nonumber\\
N_{1,1;3}(f)&=&\frac{24(N-1)}{35}\,f^5+O(f^7).
\eea
The perturbation expansion for $N_{1,2;l}(f)$ to fifth
order in $f$ is given by
\bea
N_{1,2;0}(f)&=&
\frac{40(N-1)^2}{21}\, f^5+O(f^7),\nonumber\\
N_{1,2;1}(f)&=&\frac43+\frac{10(N-1)}9\,f^2-
\left(\frac{40(N-1)^2}{27}-\frac{598(N-1)}{63}\right)\,f^4+
O(f^6),\nonumber\\
N_{1,2;2}(f)&=&-4(N-1)\,f^3+
\left(\frac{34(N-1)^2}3-\frac{404(N-1)}7\right)\,f^5+O(f^7),
\nonumber\\
N_{1,2;3}(f)&=&1-\frac{4(N-1)}{21}\,f^4+O(f^6),
\nonumber\\
N_{1,2;4}(f)&=&\frac{48(N-1)}{35}\,f^5+O(f^7).
\eea
All other fusion rules are zero to fifth order. We observe
that orthogonality is violated to fifth order for the first
and second observable.

With the fusion rules we can compute correlation functions of our
observables. Let us consider for example a general two point
function. It depends on the hierarchical distance $k$ and the
hierarchical lattice size $k^\prime$ of the system. The explicit
formula is
\beq
\left\langle{\cal O}_n(\phi|f){\cal O}_m(\phi|f)\right\rangle
=\prod_{j=1}^k\left(\epsilon_n\left(\beta^j(f)\right)
\epsilon_m\left(\beta^j(f)\right)\right)
\sum_{l=0}^\infty N_{n,m;l}\left(\beta^k(f)\right)
\prod_{i=1}^{k^\prime-k} \epsilon_l
\left(\beta^{k+i}(f)\right)
\label{two}
\eeq
for the two point function on the point of the renormalized
trajectory parametrized by $f$. This formula simply says that
each observable is renormalized independently $k$-times.
Each renormalization step produces a factor given by the
eigenvalue of the corresponding observable at the location
on the renormalized trajectory. After $k$ steps the observables
land in the same block and are there fused together. The result
of fusion is then renormalized $(k^\prime-k)$-times to
obtain the value of the two point function. The thermodynamic
limit corresponds to $k^\prime=\infty$. The formula
(\ref{two}) is still true in this limit provided that one changes
to a different parametrization of the renormalized trajectory
at the point where the running coupling $f$ diverges.

\section{Numerical results}
When dealing with perturbation theory it is natural to question
its validity. To tackle this problem we have determined our
perfect observables and their corresponding eigenvalues numerically.
The main technical task is to compute the transformations
\begin{eqnarray}
{{\cal R}V(\psi)} &=&-2\ln\left(
\int{\rm d}\mu_\gamma(\zeta)
e^{-V(\psi+\zeta)}\right), \label{RV}\\
{\cal L}_V{\cal RO}(\psi)&=&
\frac{\int{\rm d}\mu_\gamma(\zeta)
e^{-V(\psi+\zeta)}{\cal O}(\psi+\zeta)}
{\int{\rm d}\mu_\gamma(\zeta)
e^{-V(\psi+\zeta)}}\label{LRO}.
\end{eqnarray}
Both equations can be reduced to integrals of the type
\begin{equation}
  {\cal I}_F(\psi) = \int{\rm d}\mu_\gamma(\zeta) F(|\psi+\zeta|)
\end{equation}
with certain scalar functions $F(\varphi)$.
A shift $\zeta \rightarrow \psi + \zeta$
in the fluctuation field yields
\begin{equation}
   {\cal I}_F(|\psi|) = (2\pi\gamma)^{N/2}
   \int{\rm d}^N\zeta \exp\left(-\frac{1}{2\gamma}[\zeta^2+\psi^2]
   +\frac{1}{\gamma}\psi\zeta \right)F(|\zeta|).
\end{equation}
By using polar coordinates and integrating out the polar angles
we find for $N=3$
\begin{equation}
   {\cal I}_F(|\psi|) = {\cal N}
   \int \limits_0^\infty {\rm d}R
   \exp\left(-\frac{1}{2\gamma}[R^2+|\psi|^2]\right)
   \sinh\left(\frac{|\psi|R}{\gamma}\right)\frac{R F(R)}{|\psi|}.
\end{equation}
In this form the integral can be evaluated by standard numerical methods.

Equipped with integrators for (\ref{RV}) and (\ref{LRO})
the strategy goes a follows:
The first step is to determine $V_{\rm RT}$. For this purpose
we start with the perturbatively improved perfect action
$V_{\rm RT}^{\rm pert}(\phi|r_0)$ at a given radius $r_0$
as bare potential and iterate the RG transformation (\ref{RV}).
In each step the resulting potential is driven closer and closer
towards the
RT. After 10 steps we end up with a good approximation of $V_{\rm RT}(r)$ at
some radius $r=r(r_0)$.
In this iterative process the potential $V(\varphi)$ is represented as a
cubic spline with $N_{\varphi}$ equidistant knots $\varphi_i$
in the range $I_{\varphi}=[\varphi_{\rm min},\varphi_{\rm max}]$.
For each iteration one has to evaluate $V'={\cal R}V$ at these points.
The interval $I_{\varphi}$ must to be chosen in such a way
that  for the computation of
$V'(\varphi \in I_{\varphi})$
contributions $V(\varphi \not\in I_{\varphi})$ are numerically
negligible in (\ref{RV}).

\begin{figure}[h]
\centerline{
\hbox{
  \epsfig{file=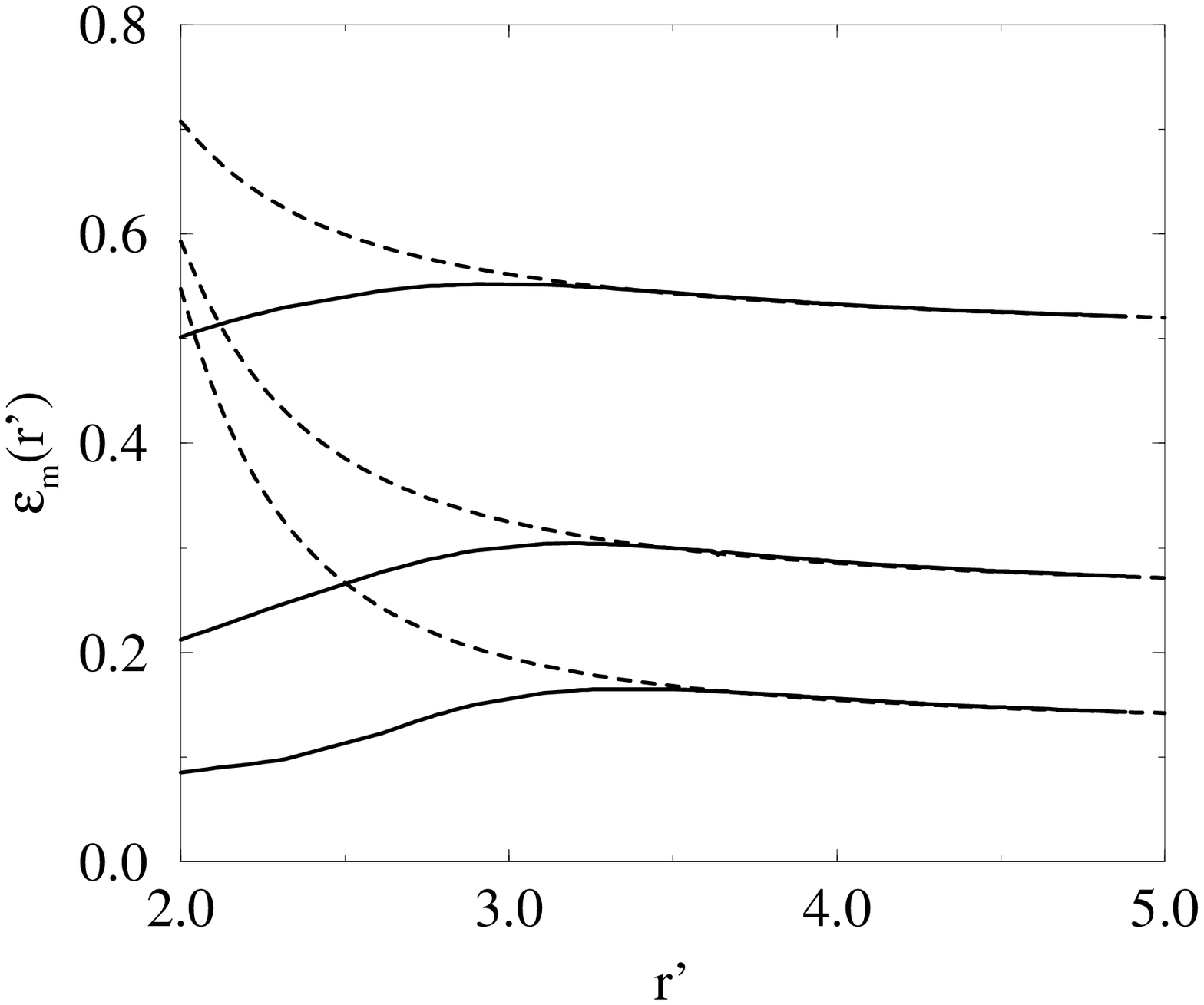,height=7.5cm,width=7.5cm,clip=}
    \epsfig{file=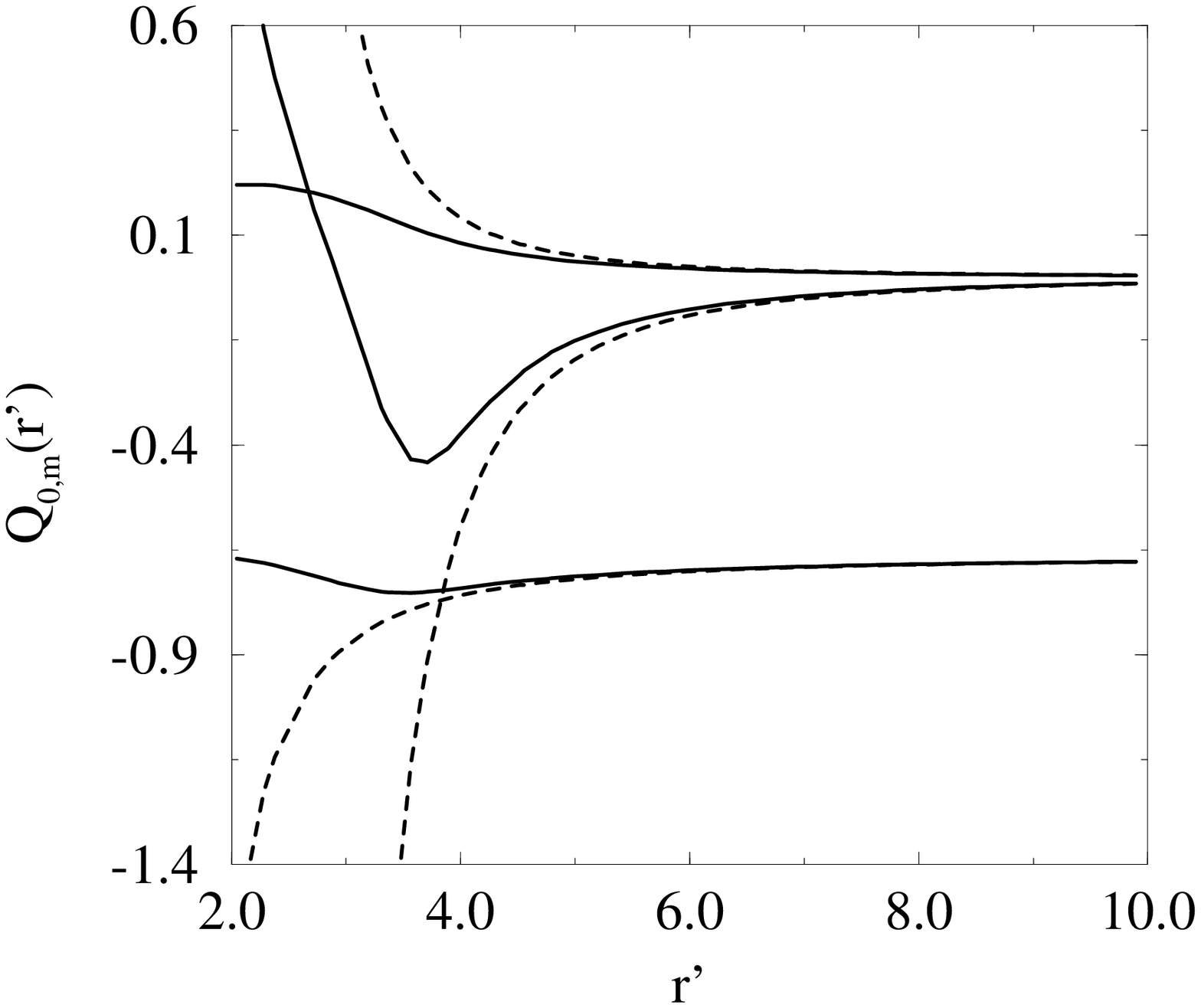,height=7.5cm,width=7.5cm,clip=}}}
\centerline{\parbox{12.5cm}{\caption[flow]{
Flow of the eigenvalues $\epsilon_m(r')$ and the
expansion coefficients $Q_{n,m}(r')$
along the renormalized trajectory.
{}From top to down $\epsilon_1,\epsilon_2,\epsilon_3$ and
$Q_{0,1},Q_{0,3},Q_{0,2}$.
Solid lines represent numerical data, dashed lines perturbative results.
Calculations were done with
$N_{\varphi}=200, \varphi_{\rm min}=0, \varphi_{\rm max}=30,M=10$.
\label{EVOBSflow}}}}
\end{figure}

In the second step the eigenvalues and eigenoperators of the linearized
RG transformation ${\cal L}_{\rm RT}{\cal R}$ at this very potential
$V_{\rm RT}(r)$ are computed.
The operator ${\cal L}_{\rm RT}{\cal R}$
acts on the infinite dimensional space of
observables. Naturally a computer can only handle the restriction
of ${\cal L}_{\rm RT}{\cal R}$ to a finite subspace.
Our program uses the space spanned by the operators
$B_m(\phi|r)=(|\phi|-r)^m$
with $m=0,\ldots,M$.
To obtain the representation matrix $L_{n,m}$ of ${\cal L}_{\rm RT}{\cal R}$
in this basis the image of every  $B_m(\phi|r)$ under
${\cal L}_{\rm RT}{\cal R}$
is numerically expanded
by a finite difference method in terms
of $B_n(\phi|r')$
\begin{equation}
  ({\cal L}_{\rm RT}{\cal R})B_m(\phi|r)
= \sum \limits_{n=0}^M L_{n,m} B_n(\phi|r').
\end{equation}
Finally  the desired eigenvectors
${\cal O}_m(r')=\sum \limits_{n=0}^{M}Q_{n,m}(r') B_n(r')$ and their
eigenvalues $\epsilon_n(r')$ are calculated from $L_{n,m}$.
Fig. \ref{EVOBSflow} shows the flow of the largest eigenvalues
$\epsilon_m(r')$.
As in the case of the potential the perturbative predictions are in
excellent agreement with the numerical results down to a radius of
about $r'\approx 4$.
Then nonperturbative effects show up forcing the eigenvalues to become
smaller.
A similar behaviour can be found for the expansions coefficients
$Q_{n,m}(r')$ except that deviations already show up at $r'\approx10$.

At $r=r_{\rm cr}\approx 2.04$ the effective radius $r'(r)$
vanishes and $r$ ceases to be an appropriate parametrization of the RT
but the eigenvalues and eigenvectors continue to flow against their
fixpoint values at the HT-fixpoint.

\begin{appendix}
\section{Hierarchical renormalization group}
In a hierarchical lattice field theory in the form
advocated by Gallavotti \cite{G} and by Gawedzki
and Kupiainen \cite{GK} the massless free covariance
$(-\bigtriangleup)^{-1}$ is replaced by the
hierarchical approximation
\beq
v(x,y)=\gamma\sum_{m=0}^n L^{m(2-D)}
\delta_{[L^{-m}x],[L^{-m}y]}.
\label{hcov}
\eeq
Its parameters are a positive real number $\gamma$
and the positive integer valued block size $L$.
The bracket $[L^{-m}x]$ denotes the integer part
of $L^{-m}x$. The lattice is $\{0,1,\dots,L^n-1\}^D$
with $n\in\N\cup\{\infty\}$. The main property of
(\ref{hcov}) is the hierarchical splitting
\beq
v=L^{2-D}C^T v C+\Gamma,
\label{split}
\eeq
$C$ being the unnormalized block average
$C(x,y)=\delta_{x,[L^{-1}y]}$ and $\Gamma$ the
diagonal part $\Gamma(x,y)=\gamma\delta_{x,y}$.
The hierarchical covariance on the right hand side
is understood to live on a lattice shrunk by
the blocking scale $L$. The splitting (\ref{split})
induces a decomposition of the lattice field $\phi$ into
a fluctuation field $\zeta$ and a background field
$\psi$. The associated block spin transformation is
\beq
e^{-{\cal R}{\cal V}(\psi)}=
{\cal N}\int {\rm d}\mu_\Gamma(\zeta)
e^{-{\cal V}(L^{1-\frac D2}C^T\psi+\zeta)}.
\label{block}
\eeq
Here ${\rm d}\mu_\Gamma(\zeta)$ is the Gaussian measure
on field space with mean zero and covariance $\Gamma$.
The $n$-fold iteration of (\ref{block}) computes the
hierarchical partition function (when the normalization
is ${\cal N}=1$). The transformation (\ref{block}) preserves
locality. For ${\cal V}(\phi)=\sum_x V(\phi(x))$ we find
${\cal R}{\cal V}(\phi)=\sum_x {\cal R}V(\phi(x))$
in a selfexplanatory notation. (${\cal R}$ is not meant to
be linear.) (\ref{block}) then reduces to the non-linear
transformation
\beq
e^{-{\cal R} V(\psi)}=
{\cal N}\left(\int {\rm d}\mu_\gamma(\zeta)
e^{-V(L^{1-\frac D2}\psi+\zeta)}\right)^{L^D}.
\label{lblock}
\eeq
on functions of a single variable $\phi$. Note that
$C^T\psi$ is constant on blocks and that $L^D$ is the
number of sites in one block. The linear block spin
transformation for observables is
\beq
{\cal L}_{\cal V}{\cal R O}(\psi)=
\frac{\int{\rm d}\mu_\Gamma(\zeta)
e^{-{\cal V}(L^{1-\frac D2}C^T\psi+\zeta)}
{\cal O}(L^{1-\frac D2}C^T\psi+\zeta)}
{\int {\rm d}\mu_\Gamma(\zeta)
e^{-{\cal V}(L^{1-\frac D2}C^T\psi+\zeta)}}.
\label{obs}
\eeq
The expectation value of an effective observable in
the effective theory equals that in the original theory.
(\ref{lobs}) preserves the property of factorization
provided that the interaction is local.
A local observable ${\cal O}(\phi(x))$ is defined as
a function of the field $\phi$ at a single site $x$.
Local observables are again transformed into local
observables according to
\beq
{\cal L}_V{\cal R O}(\psi)=
\frac{\int{\rm d}\mu_\gamma(\zeta)
e^{-V(L^{1-\frac D2}\psi+\zeta)}
{\cal O}(L^{1-\frac D2}\psi+\zeta)}
{\int {\rm d}\mu_\gamma(\zeta)
e^{-V(L^{1-\frac D2}\psi+\zeta)}}.
\label{lobs}
\eeq
The transformation (\ref{obs}) is the linearization of
(\ref{block}) at the interaction ${\cal V}(\phi)$.
The local transformation (\ref{lobs}) is the linearization
of (\ref{lblock}) at the local interaction $V(\phi)$
divided by the block volume $L^D$.
The expectation value of a local observable is the
$n$-fold iteration of (\ref{lobs}) evaluated at
zero external field. The thermodynamic limit
$n=\infty$ follows from an infinite iteration.

The computation of higher correlation functions requires
another ingredience called fusion. Let us consider for
instance a bilocal observable of the product form
${\cal O}_1(\phi(x_1)){\cal O}_2(\phi(x_2))$ depending on the
field $\phi$ at two sites $x_1$ and $x_2$.
The hierarchical distance of $x_1$ and $x_2$ is given
by ${\rm dist}(x_1,x_2)=L^{{\rm k}(x_1,x_2)}$, where
${\rm k}(x_1,x_2)$ is the smallest positive integer such that
$[L^{-{\rm k}(x_1,x_2)}x_1]=[L^{-{\rm k}(x_1,x_2)}x_1]$.
Both observable are then transformed independently
${\rm k}(x_1,x_2)$-times according to the local transformation
(\ref{lobs}) until $x_1$ and $x_2$ land in the
same block, whereupon the effective observables are multiplied
together. We call this process fusion. The result is then
again transformed $n-{\rm k}(x_1,x_2)$-times according to the
local rule (\ref{lobs}) to give the hierarchical
two point function. In the thermodynamic limit $n=\infty$
the observable has to be renormalized an infinite
number of steps after fusion. The process can be visualized
as a fork.

Both (\ref{lblock}) and (\ref{lobs}) make sense when the
dimension parameter $D$ and the scale parameter $L$ are
continued to real values.
\end{appendix}

\begin{thebibliography}{XXX}
%
\bibitem[HN]{HN} P. Hasenfratz and F. Niedermayer,
Perfect lattice action for asymptotically free theories,
Nucl. Phys. B414 (1994) 785-814
%
\bibitem[G]{G} G. Gallavotti,
Some aspects of the renormalization problems in
statistical mechanics and field theory,
Mem. Accad. Lincei 15, 23 (1978)
%
\bibitem[GK]{GK} K. Gawedzki and A. Kupiainen,
1) Asymptotic freedom beyond perturbation theory,
HUTMP 85/B177 (1985),
2) Continuum limit of the hierarchical $O(N)$ non-linear
$\sigma$-model,
Comm. Math. Phys. 106, 533-550 (1986),
3) Nontrivial continuum limit of a $\phi_4^4$-model with
negative coupling constant,
Nucl. Phys. B257, FS14, 474-504 (1985)
%
\bibitem[PR]{PR} A. Pordt and T. Reisz,
On the renormalization group iteration of a two dimensional
hierarchical non-linear $O(N)$ $\sigma$-model,
BI-TP 89/40
%
\bibitem[S]{S} K. Symanzik,
Continuum limit and improved action in lattice theories,
(I) Principles and $\phi^4$ theory,
Nucl. Phys. B226 (1983) 187-204;
(II) $O(N)$ non-linear $\sigma$ model in perturbation theory,
Nucl. Phys. B226 (1983) 205-227
%
\bibitem[WK]{WK} K. Wilson and J. Kogut,
The renormalization group and the $\epsilon$ expansion,
Phys. Rep. C12, No. 2 (1974) 75-200
%
\bibitem[WX]{WX} C. Wieczerkowski and Y. Xylander,
Improved actions, the perfect action, and scaling by
perturbation theory in Wilsons renormalization group:
the two dimensional $O(N)$-invariant non-linear
$\sigma$-model in the hierarchical approximation,
Nucl. Phys. B440 (1995) 393
\end{thebibliography}
\end{document}